\documentclass{elsart}

\usepackage{amssymb}

\begin{document}

\begin{frontmatter}



\title{On connection between purity and fidelity of quantum channels}


\author[label1,label2]{A.V. Kuzmin}

\address[label1]{Joint Institute for Power and Nuclear Research, NAS of Belarus, 220109 Minsk,
Krasina str. 99, Belarus}
\address[label2]{Center for Applied Mathematics and Theoretical Physics, University of Maribor, Krekova 2,
SI-2000 Maribor, Slovenia, European Union}

\begin{abstract}
Quantum channels depending on a number of classical control parameters are considered. Assuming the stochastic
fluctuations of the control parameters in the small errors limit it is shown that the channel fidelity is equal to the 
average value of the channel purities calculated in the cases of the control errors presence and absence respectively. 
The result validity is demonstrated on two particular examples, namely, one-qubit quantum gates implemented in the framework of the
ion trap setup and the single qubit anisotropic depolarizing quantum channel. 
\end{abstract}

\end{frontmatter}

\section{Introduction}

The investigation of open quantum systems is of importance in various fields, for example, in  quantum information 
science. When we consider a quantum system interacting with a classical environment, the corresponding linear operators 
(observables) depend on a set of the classical parameters representing the particular configuration of the environment.
The decoherence phenomenon appearing during the evolution of open quantum system as well as the quantum dynamics stability 
are of interest especially in the context of quantum information
processing and, more generally, foundations of quantum physics. The formalism of quantum channels is very useful in describing 
quantum dynamics, including decoherence phenomenon~\cite{Kraus}. Thus, in order to phenomenologically  describe both the coherent 
(unitary evolution) and 
non-coherent (decoherence) dynamics of quantum system in a classical environment  one can exploit the formalism of quantum channels 
depending on a number of the classical {\it control} parameters. While one is interested in a quantum evolution of the single (open) 
quantum system and takes into account its interaction with the environment phenomenologically the decoherence strength can be 
quantified 
by comparing the system's final state purity (see~\cite{Zurek91}) with the initial state one. The generally accepted measure of 
quantum evolution stability is the fidelity~\cite{Peres}, for the recent developments in this field, for instance, 
see~\cite{ProsenJPA2003}. The question about the connection between the purity and the fidelity of quantum channels was touched in a
number of papers. In the Ref.~\cite{DecLE2003} the relation between the temporal behavior of the purity and the average Loschmidt echo 
(fidelity) was investigated on a rather general example of Hamiltonian system. It was demonstrated that the squared average 
Loschmidt echo is less or equal to the state purity taken 
at the same time moment. Fidelity and purity decay in weakly coupled composite systems was studied in the Ref.~\cite{ProsenJPA2003}.
The general inequality $F^2 \le F_R^2 \le I$ was reported, see also~\cite{ProsenPRA2003}. Here $F$ is the fidelity, $F_R$ is 
the reduced fidelity and $I$ is the purity. All quantities are taken at the same time moment. For rigorous definitions 
see~\cite{ProsenJPA2003}. The pure-state fidelity and the purity of quantum channels was studied in the Ref.~\cite{ZanardiLidar2004}. 
The channel purity and fidelity Hamiltonians were introduced which expectation values give either the channel purity or 
pure-state fidelity. Some
particular channel examples were considered. Stability and instability of quantum evolution were studied in the interaction 
between a two-level atom with photon recoil and a quantized field mode in an ideal cavity in the Ref.~\cite{Rus2005}. Particularly,
it was shown both with initial Fock and coherent field states that quantum entropy, purity and fidelity of regularly moving atoms 
evolve in a regular way, whereas the respective quantum evolution is unstable with atoms moving chaotically. The transition of quantum
system from a pure state to a mixed one, which is induced by the quantum criticality of the surrounding system modeled as an Ising 
spin chain in a transverse field was investigated in the Ref.~\cite{ZanardiPRL2006}. Particularly interesting in the present context,
the linear relationship between the purity of the central system's reduced density matrix and Loschmidt echo was obtained in the 
framework of the model 
considered. The similar linear formula connecting the final state purity and the Hadamard  gate fidelity was derived in the context of
optical and ion trap holonomic quantum computations with small stochastic squeezing control errors~\cite{wePRA2006}.

In this Letter the quantum channels depending on a number of stochastically fluctuating classical control parameters are considered. 
The linear relationship between the channel purity and fidelity is obtained in the small control errors limit. Namely, it is 
demonstrated that the channel fidelity equals the average value of the channel purities calculated in the cases of the control 
errors presence and absence respectively. The validity of this 
relation is checked for the two particular implementations of quantum channels, namely, arbitrary one-qubit quantum gates 
implemented in the framework of the ion trap setup~\cite{2levelTRIO} and the single qubit anisotropic depolarizing channel.      

\section{The relationship between channel purity and fidelity in small control errors limit}

The channel $T$ over the finite-dimensional quantum state space ${\it H}$ is a completely
positive trace-preserving quantum map, which has a non unique Kraus operator-sum representation~\cite{Kraus}:
\begin{equation}\label{Kraus}
  T(\rho, \lambda) = A_i \left( \lambda \right)  \rho A_i^\dag \left( \lambda  \right), 
\end{equation}
where the Kraus operators $A_i$ satisfy the constraint $A_i A_i^\dag = 1$ (here and bellow the summation over the repeated 
indexes is assumed, exceptions are evident) and $\rho$ is the initial state density matrix. In this work it is also assumed 
that the channel $T$ depends 
on a number of the classical control parameters represented by the vector $\lambda \equiv \left\{ \lambda_\mu  \right\}_{\mu=1}^N$, 
where the index $\mu$ runs from $1$ to $N$. Let the channel $T\left( \rho , \lambda  \right)$ to be implemented in the ideal 
case of control errors absence. In reality the control errors (fluctuations of the control parameters) are unavoidable. Thus instead
of the ideal channel implementation $T\left( \rho , \lambda  \right)$  one gets $T\left( \rho , \lambda + \delta \lambda \right)$. 
Here the only one source of errors is considered, namely, the errors in the assignment of the classical control parameters. The 
quantity $\delta \lambda$ represents the random fluctuations of the parameters. According to the rules of quantum mechanics one has 
to average the channel $T\left( \rho , \lambda + \delta \lambda \right)$ over the fluctuations in order to obtain the final state 
density matrix.       

Let us denote the purity of the final state obtained when the initial state $\rho$ is transmitted through the channel in the case of 
no errors and when the parameters values are $\lambda$ as $P_0 (\rho , \lambda) = tr\left[ T\left( \rho , \lambda \right) \right]$. 
The purity of the channel $T$, when the initial state density matrix equals $\rho$, in the case of stochastic control errors is given 
by the expression:
\begin{equation}\label{ChPurity}
  P(\rho, \lambda) = tr\left[ \overline{T\left( \rho, \lambda + \delta \lambda \right)}^2 \right].
\end{equation}
Here and bellow the line over the quantity denotes the averaging procedure over the control parameters fluctuations. 
The channel fidelity for the initial state described by $\rho$ and parameters values $\lambda$ is 
\begin{equation}\label{ChFidelity}
  F\left( \rho , \lambda  \right) = tr\left[ T\left( \rho , \lambda  \right) \overline{T\left( \rho , \lambda + 
\delta \lambda \right)} \right].
\end{equation}
One has to distinguish two cases in the small control errors limit. The first one when $\overline{\delta \lambda_\mu} \ne 0$, and 
the second one when $\overline{\delta \lambda_\mu} = 0$, both for all $\mu$. Consider the first case of non-zero average values of 
control errors. For the 
channel purity when the initial state is described by $\rho$ one gets:
\begin{equation}\label{Pn0}
  P \left( \rho , \lambda \right) \left. \right|_{\overline{\delta 
\lambda_\mu} \ne 0} = P_0 (\rho , \lambda)  + 2 tr \left( T \partial_\mu T  \right) \overline{\delta \lambda_\mu} + \ldots .
\end{equation}  
Here the leading order terms are remained only and $T \equiv T \left( \rho , \lambda \right)$. The channel fidelity under the same
assumptions is
\begin{equation}\label{Fn0}
  F \left( \rho , \lambda \right)\left. \right|_{\overline{\delta 
\lambda_\mu} \ne 0} = P_0 (\rho , \lambda) + tr \left( T \partial_\mu T \right) \overline{\delta \lambda_\mu} + \ldots .
\end{equation}
From the Eqs.~(\ref{Pn0}) and (\ref{Fn0}) it is immediately seen that the following relationship is held:
\begin{equation}\label{fin0}
  F \left( \rho , \lambda \right)\left. \right|_{\overline{
\delta \lambda_\mu} \ne 0}= \frac{P \left( \rho , \lambda \right) \left. \right|_{\overline{
\delta \lambda_\mu} \ne 0}+P_0 (\rho , \lambda)}{2}.
\end{equation} 
It means that the channel fidelity equals the average value of the channel purities calculated in the cases of the stochastic 
control errors presence and absence respectively. 

Now we assume that the average values of the control errors equal zero, namely $\overline{\delta \lambda_\mu} = 0$ for all $\mu$. 
Under this assumption in the small control errors limit one gets: 
\begin{eqnarray}\label{P}
    P (\rho , \lambda)\left. \right|_{\overline{\delta 
\lambda_\mu} = 0} = P_0 (\rho , \lambda)  + tr \left( T \partial_\mu \partial_\nu T \right) \overline{\delta \lambda_\mu \delta \lambda_\nu}+
\ldots .
\end{eqnarray}
In the last expression we have neglected the higher order terms
that is justified by the usage of the small control errors limit as mentioned above. Under
the same assumptions the channel fidelity has the following form:
\begin{eqnarray}\label{F}
    F (\rho , \lambda)\left. \right|_{\overline{\delta 
\lambda_\mu} = 0} = P_0 (\rho , \lambda) + \frac{1}{2} tr \left( T \partial_\mu \partial_\nu T \right) \overline{\delta \lambda_\mu 
\delta \lambda_\nu} + \ldots.
\end{eqnarray}
Comparing the Eqs.~(\ref{P}) and (\ref{F}) as well taking into account the Eq.~(\ref{fin0}) we can conclude that in
the small errors limit independently whether the average value of the control errors equals zero or not the following relationship 
between the channel fidelity and purity is held:
\begin{equation}\label{IP}
    F (\rho , \lambda) = \frac{P (\rho , \lambda) + P_0 (\rho , \lambda)}{2},
\end{equation}
Thus the channel fidelity equals the average value of the channel purities calculated in the cases of the control errors presence 
and absence respectively. The expression~(\ref{IP}) was derived in the case of the particular Hadamard gate implementation on
optical and ion trap holonomic quantum computers in the Ref.~\cite{wePRA2006}. Similar relationship between the purity and Loschmidt 
echo was obtained in the Ref.~\cite{ZanardiPRL2006} in the context of the central system interacting with the surrounding system 
modeled by the Ising spin chain in a transverse field.    

\section{Example 1: one-qubit ion trap quantum computations}

The relationship~(\ref{IP}) must be preserved in the particular case when the quantum channel represents some quantum gate, 
i.e. unitary
evolution. In this case Kraus operator-sum representation of the channel  reduces to $T \left( \rho , \lambda  \right) = 
U\left( \lambda \right) \rho U^\dag \left( \lambda \right)$, where $U \left( \lambda \right)$ is a unitary operator. In the 
Ref.~\cite{wePRA2006} in the particular case of Hadamard gate
implemented on optical and ion trap holonomic quantum computers it was obtained that the relationship~(\ref{IP}) is satisfied.
Now one more evidence supporting the validity of the Eq.~(\ref{IP}) will be given. Namely, it will be derived in the case of
the arbitrary one-qubit gate implemented on the ion trap quantum computer setup proposed in the Ref.~\cite{2levelTRIO}. 
In the framework of this approach arbitrary one-qubit gate is given by the expression:
\begin{equation}
  R \left( \theta , \varphi  \right) = \exp{i\frac{\theta}{2}\left( e^{i\varphi} \sigma^+ + e^{- i \varphi} \sigma^-  \right)},
\end{equation}
where $\sigma^{\pm} = \left( \sigma_x \pm i\sigma_y \right)/2$, quantities $\theta$ and $\varphi$ are the classical control 
parameters and $\sigma_{x,y}$ denote Pauli matrices. According to the general theory developed in the previous section assume that
the control parameters stochastically fluctuate near the desired values $\theta_0$ and $\varphi_0$. Namely, $\theta = \theta_0 +
\delta \theta$ and $\varphi = \varphi_0 + \delta \varphi$. As well fluctuations obey the condition $\overline{\delta \theta} = 
\overline{\delta \varphi} = 0$. Here the line denotes averaged quantities. After a bit lengthy but straightforward calculations one 
obtains the following expressions for the channel (final state) purity and channel (quantum gate) fidelity:
\begin{eqnarray}\label{QC:PF}
  P (\rho , \theta_0 , \varphi_0) = tr \rho^2 + 2 tr G_0 \overline{G}_2, \nonumber \\
  F (\rho , \theta_0 , \varphi_0) = tr \rho^2 + tr G_0 \overline{G}_2,
\end{eqnarray}   
where
\begin{eqnarray}\label{Gs}
  G_0 = &&h_0 + \frac{i}{2} f_0 \left[ x_0 , \rho \right] + \frac{1}{2} g_0  x_0 \rho x_0 , \nonumber \\
  G_2 = &&h_2 + \frac{i}{2} f_2 \left[ x_0 , \rho \right] + \frac{i}{2} f_1 \left[ x_1 , \rho \right] + \frac{i}{2} f_0
\left[ x_2 , \rho \right] \nonumber \\
&&+ \frac{1}{2} g_0 x_0 \rho x_2 + \frac{1}{2} g_0 x_2 \rho x_0 + \frac{1}{2} g_1 x_1 \rho x_0 \nonumber \\
&&+ \frac{1}{2} g_1 x_0 \rho x_1 + \frac{1}{2} g_2 x_0 \rho x_0 + \frac{1}{2} g_0 x_1 \rho x_1,   
\end{eqnarray}
and 
\begin{eqnarray}\label{hgx}
  h_0 = \frac{1}{2} \rho \left( 1+ \cos{\theta_0} \right), \nonumber \\
  h_2 = - \frac{1}{4} \rho \cos{\theta_0} \delta \theta^2 , \nonumber \\
  f_0 = \theta_0^{-1} \sin{\theta_0}, \nonumber \\
  f_1 = \left( \theta_0^{-1} \cos{\theta_0} - \theta_0^{-2} \sin{\theta_0} \right) \delta \theta , \nonumber \\
  f_2 = \frac{1}{2} \left( 2 \theta_0^{-3} \sin{\theta_0} - 2\theta_0^{-2} \cos{\theta_0} - \theta_0^{-1} \sin{\theta_0} \right)
  \delta \theta^2 , \nonumber \\
  g_0 = \theta_0^{-2} \left( 1 - \cos{\theta_0} \right) , \nonumber \\
  g_1 = \left( \theta_0^{-2} \sin{\theta_0} - 2 \theta_0^{-3} \left( 1 - \cos{\theta_0} \right) \right) \delta \theta ,
  \nonumber \\
  g_2 = \frac{1}{2} \left( 6\theta_0^{-4} \left( 1 - \cos{\theta_0} \right) - 4 \theta_0^{-3} \sin{\theta_0} + \theta_0^{-2}
  \cos{\theta_0}  \right) \delta \theta^2 , \nonumber \\
  x_0 = \theta_0 \left( e^{i \varphi_0} \sigma^+ + e^{-i \varphi_0} \sigma^-  \right), \nonumber \\
  x_1 = i \theta_0  \left( e^{i \varphi_0} \sigma^+ - e^{-i \varphi_0} \sigma^-  \right)\delta \varphi +
   \left( e^{i \varphi_0} \sigma^+ + e^{-i \varphi_0} \sigma^-  \right)\delta \theta , \nonumber \\
  x_2 = - \frac{1}{2} \theta_0  \left( e^{i \varphi_0} \sigma^+ + e^{-i \varphi_0} \sigma^-  \right) \delta \varphi^2 +
  i  \left( e^{i \varphi_0} \sigma^+ - e^{-i \varphi_0} \sigma^-  \right) \delta \varphi \delta \theta .
\end{eqnarray}
From the Eqs.~(\ref{QC:PF}) it immediately follows that 
\begin{equation}\label{QC:Fin}
  F (\rho , \theta_0 , \varphi_0) = \frac{P (\rho , \theta_0 , \varphi_0) + P_0 (\rho)}{2},
\end{equation}
where $P_0 (\rho) = tr \rho^2$. Thus in the particular case of one-qubit quantum gates implemented on the ion trap quantum computer 
the fidelity of the arbitrary gate equals the average value of the final state purities calculated in the cases of the control 
errors presence and absence respectively. In the later case the final state purity $P_0 (\rho)$ coincides with the initial state 
one since the evolution is unitary.

\section{Example 2: single qubit anisotropic depolarizing channel}

Consider the single qubit anisotropic depolarizing channel defined by the following Kraus operator-sum representation:
\begin{equation}\label{ADC}
  T_{ADC} \left( \rho , \left\{ p_i \right\}_{i=0}^{3} \right) = \sum\limits_{i=0}^{3} p_i \sigma_i \rho \sigma_i ,
\end{equation} 
where $\sigma_i$ with $i=1,2,3$ are Pauli matrices, $\sigma_0$ equals the identity matrix and $\left\{ p_i \right\}_{i=0}^{3}$ is a
probability distribution playing the role of the control parameters vector. At the channel output the averaged final state density
matrix is given by the expression:
\begin{equation}\label{ADCrho}
  \overline{T_{ADC} \left( \rho , \left\{ p_i + \delta p_i \right\}_{i=0}^{3} \right)} = 
\sum\limits_{i=0}^{3} \left( p_i + \overline{\delta p_i} \right) \sigma_i \rho \sigma_i.
\end{equation} 
It is seen that one has to accept that $\overline{\delta p_i} \ne 0$ at least for one $i$ overwise the case is trivial. One can assume
that $\overline{\delta p_i} \ne 0$ for all values of the index $i$ without loosing the generality. By means of the straightforward 
calculations it is easy to demonstrate that the channel purity for the initial state $\rho$ is
\begin{equation}\label{Padc}
  P_{ADC} \left( \rho , \left\{ p_i \right\}_{i=0}^3 \right) = P_{ADC}^{(0)} + 2 tr \left( \sum\limits_{i,j=0}^3 
p_i \overline{\delta p_j} \sigma_i \rho \sigma_i \sigma_j \rho \sigma_j \right).
\end{equation}
Here $P_{ADC}^{(0)}$ is the final state purity at the channel output in the case of the control errors absence. Similarly, one gets
the expression for the fidelity:
\begin{equation}\label{Fadc}
  F_{ADC} \left(  \rho , \left\{ p_i \right\}_{i=0}^3  \right) = P_{ADC}^{(0)} + tr \left( \sum\limits_{i,j=0}^3 
p_i \overline{\delta p_j} \sigma_i \rho \sigma_i \sigma_j \rho \sigma_j \right).
\end{equation} 
Comparing the Eqs.~(\ref{Padc}) and (\ref{Fadc}) it is evident that the relationship~(\ref{IP}) is held.

\section{Conclusions}

Quantum channels depending on a number of classical control parameters were considered. Assuming the stochastic
fluctuations of the parameters in the small errors limit it was shown that the channel fidelity is equal to the 
average value of the channel purities calculated in the cases of the control errors presence and absence respectively. 
Thus, the close relationship between the stability of quantum evolution 
and the decoherence induced by stochastic control errors is demonstrated. The result validity was checked on two particular 
examples, namely, the one-qubit ion trap quantum gates and the single qubit anisotropic depolarizing quantum channel. 
The advantage of the relationship obtained is that in the case when one investigates the decoherence
induced by stochastic control errors or the stability of quantum dynamics it is not necessary to calculate both the channel (final 
state) purity and the channel (quantum gate) fidelity. In the small errors limit it is enough to calculate 
one of these quantities that can significantly reduce the size of the work to be done. As well the simple relationship obtained 
gives a good starting point for the estimates when the control errors are not small. Particularly, the presented result can be used
when one investigates the stability of holonomic quantum computations with respect to control errors, see~\cite{wePRA2006,wePLA} and
references therein.

\section{Acknowledgements}

I would like to thank Prof. Marko Robnik for his warm hospitality at CAMTP, University of Maribor. 
The financial support by Ad Futura foundation is gratefully acknowledged.


\begin{thebibliography}{00}




\bibitem{Kraus} K. Kraus, {\it States, Effects and Operations, Fundamental Notions of Quantum Theory} (Academic, Berlin, 1983).

\bibitem{Zurek91} W.H. Zurek, Phys.\/Today 44 (1991) 36.

\bibitem{Peres} A. Peres, Phys.\/Rev.\/A 30 (1984) 1610.

\bibitem{ProsenJPA2003} M. Znidarich and T. Prosen, J.\/Phys.\/A 36 (2003) 2463.

\bibitem{DecLE2003} F.M. Cucchietti {\it et al.}, Phys.\/Rev.\/Lett. 91 (2003) 210403.

\bibitem{ProsenPRA2003} T. Prosen, T.H. Seligman and M.~Znidaric, Phys.\/Rev.\/A 67 (2003) 042112.

\bibitem{ZanardiLidar2004} P. Zanardi and D.A. Lidar, Phys.\/Rev.\/A 70 (2004) 012315. 

\bibitem{Rus2005} S.V. Prants, M.Yu. Uleysky and V.Yu.~Argonov, e-print: quant-ph/0511106.

\bibitem{ZanardiPRL2006} H.T. Quan, {\it et al.}, Phys.\/Rev.\/Lett. 96 (2006) 140604.

\bibitem{wePRA2006} V.I. Kuvshinov and A.V. Kuzmin, Phys.\/Rev.\/A 73 (2006) 052305; quant-ph/0603282.

\bibitem{2levelTRIO} A.M. Childs and I.L. Chuang, Phys. Rev. A 63 (2001) 012306.

\bibitem{wePLA} V.I. Kuvshinov and A.V. Kuzmin, Phys.\/Lett.\/A 316 (2003) 391; Phys.\/Lett.\/A 341 (2005) 450;
quant-ph/0310069; quant-ph/0503226.

\end{thebibliography}
\end{document}